\documentclass[12pt]{article}
\usepackage[english,german,french,polish]{babel}
\usepackage[T1]{fontenc}
\usepackage{amsfonts}

\begin{document}

\selectlanguage{english}

\baselineskip 0.73cm
\topmargin -0.4in
\oddsidemargin -0.1in

\let\ni=\noindent

\renewcommand{\thefootnote}{\fnsymbol{footnote}}

\newcommand{\SM}{Standard Model }

\pagestyle {plain}

\setcounter{page}{1}



~~~~~~
\pagestyle{empty}

\begin{flushright}
IFT-- 08/8
\end{flushright}

\vspace{0.4cm}

{\large\centerline{\bf On Maxwell's equations unified with the dynamics}}

{\large\centerline{\bf of cold dark matter}}

\vspace{0.5cm}

{\centerline {\sc Wojciech Kr\'{o}likowski}}

\vspace{0.3cm}

{\centerline {\it Institute of Theoretical Physics, University of Warsaw}}

{\centerline {\it Ho\.{z}a 69, 00--681 Warszawa, ~Poland}}

\vspace{0.6cm}

{\centerline{\bf Abstract}}

\vspace{0.2cm}

\begin{small}

\begin{quotation}

New simple field equations are proposed (Eqs. (5)), unifying Maxwell's equations of the Standard Model 
electrodynamics (after the electroweak symmetry is spontaneously broken) with the dynamics of cold dark
matter. The cold dark matter is represented here by sterile spin-1/2 Dirac fermions (sterinos) whose mass 
is spontaneously generated by the nonzero vacuum expectation value of a field of sterile scalars (sterons), 
while their interactions are mediated by quanta of a sterile antisymmetric tensor field with a large mass. 
The new field equations include a specific weak coupling  of the Standard Model electromagnetic field to 
the sterile fields. It has a spontaneous quasi-magnetic structure. Such a coupling opens a photonic portal between the Standard Model world and the 
sterile world.
 
\vspace{0.6cm}

\ni PACS numbers: 14.80.-j , 04.50.+h , 95.35.+d 

\vspace{0.3cm}

\ni June 2008

\end{quotation}
 
\end{small}

\vfill\eject

\pagestyle {plain}

\setcounter{page}{1}

\vspace{0.3cm}

\ni {\bf 1. Introduction}

\vspace{0.3cm} 

The quantum electrodynamics is historically the oldest part of the structure of \SM within which it has been unified with weak interactions and correlated with strong interactions by means of the $SU(3)\times SU(2)\times U(1)$ symmetry, spontaneously broken with the use of nonzero vacuum expectation value of the neutral component of Higgs weak doublet. After the chiral electroweak symmetry $SU(2)\times U(1)$ is spontaneously broken, the electromagnetic field appears and photons emerge as physical objects. In this note, we put forward the bold suggestion that in Nature, after the spontaneous break-down of electroweak symmetry, the \SM electrodynamics becomes unified with the dynamics of cold dark matter into a new dynamical structure. Somehow, the new unification is dual or complementary to the previous electroweak unification.

Perhaps, such an idea, being not precise, may be realized in different ways. In a recent work [1], we realize it {\it de facto} by means of the conjecture that the cold dark matter, though assumed to be sterile from the \SM gauge charges, can communicate with \SM world not only through gravity, but also through a {\it photonic portal} to the sterile world. This portal is provided by a weak coupling of the electromagnetic field $F_{\mu\,\nu} = \partial_\mu A_\nu - \partial_\nu A_\mu $ to the fields responsible for cold dark matter (for an alternative Higgs portal to the sterile world see {\it e.g.} Ref. [2]){\footnote {In the possible communication between two worlds, the pronounced role of photons is due to the fact that --- after the spontaneous breaking of electroweak symmetry $SU(2)\times U(1)$ --- they are gauge bosons of an only existing unbroken symmetry (beside $SU(3)$), the relic electromagnetic symmetry $U_{\rm EM}(1)$). We believe that this fact provides a good starting point for the argumentation in favour of the existence of photonic portal between both worlds, if --- of course --- they can communicate not only through gravity.}}. In our work, we assume that the cold dark matter consists of sterile spin-1/2 Dirac fermions ({\it sterinos}), whose mass is spontaneously generated by the nonzero vacuum expectation value of a field of sterile scalars ({\it sterons}). Then, a simple realization of the photonic portal can be given by the following effective interaction Lagrangian: 

\begin{equation}
-\frac{1}{4} \frac{f}{M^2}\left[F_{\mu\,\nu}\,\varphi F^{\mu\,\nu}\,\varphi + 2\zeta(\bar\psi \sigma_{\mu\,\nu} \psi )\, F^{\mu\,\nu}\varphi \right]\;,
\end{equation}

\ni where $\psi$ and $\varphi $ are the sterino and steron fields, respectively,  $M$ denotes a large mass scale and $f>0$ as well as $2 f\zeta$ stand for unknown dimensionless coupling constants. If a third term:

\begin{equation}
-\frac{1}{4} \frac{f\zeta^2}{M^2} (\bar\psi \sigma_{\mu\,\nu} \psi ) (\bar\psi \sigma^{\mu\,\nu} \psi )
\end{equation}


\ni is added to the interaction (1), then the effective interaction Lagrangian gets the following quadratic form:

\begin{equation}
-\frac{1}{4} \frac{f}{M^2}\left(F_{\mu\,\nu}\,\varphi + \zeta \bar\psi \sigma_{\mu\,\nu} \psi \right) \left( F^{\mu\,\nu}\,\varphi + \zeta \bar\psi \sigma^{\mu\,\nu} \psi \right)\;.
\end{equation}

\ni Here, we have $\varphi\equiv  <\!\!\varphi\!\!>_{\rm vac} + \,\varphi_{\rm ph} $, while $<\!\!\varphi\!\!>_{\rm vac} \neq 0$ and $\varphi_{\rm ph}$ denotes the physical steron field. The sterino mass is spontaneously generated by $<\!\!\varphi\!\!>_{\rm vac} \neq 0$.

At a more fundamental level, such an interaction may suggest the existence of a mediating sterile field which --- in this case --- is an antisymmetric tensor field $A_{\mu\,\nu}$ (of dimension one, while $F_{\mu\,\nu}$ is of dimension two). Then, the large mass scale $M$ becomes the mass of the related particles (quanta of the field $A_{\mu\,\nu}$). They have spin 1 and parity $-$ or + , and can be described by three-dimensional vector or axial vector fields $A_{k o}$ or $\varepsilon_{k l m}A^{l m}$, respectively. The effective interaction (3) follows approximately, when momentum transfers through the field $A_{\mu\,\nu}$ are negligible with respect to the large mass $M$. Note that the mass $M$ may be also spontaneously generated by the nonzero vacuum expectation value $<\!\!\varphi\!\!>_{\rm vac} \neq 0$.

\vspace{0.3cm}

\ni {\bf 2. Extended quantum electrodynamics}

\vspace{0.3cm} 

The \SM electromagnetic Lagrangian $ -\frac{1}{4}F_{\mu\,\nu} F^{\mu\,\nu} - j_\mu A^\mu$ is suggested, therefore, to be extended by adding to it the universal interaction of the mediating sterile field $A_{\mu\,\nu}$ with an antisymmetric-tensor bilinear form (the "tensor current") built up of $\psi$, $\varphi $  and $F_{\mu\,\nu}$: $\sqrt{f}(F_{\mu\,\nu} \varphi + \zeta \bar\psi \sigma_{\mu\,\nu} \psi)$. Then, the supplemented electromagnetic Lagrangian takes the form 

\begin{eqnarray}
{\cal{ L }}\! & \!\!=\!\! & \!-\frac{1}{4}F_{\mu\,\nu} F^{\mu\,\nu} - j_\mu A^\mu - \frac{1}{4} \left[\left(\partial_\lambda A_{\mu \nu}\right)\left(\partial^\lambda A^{\mu \nu}\right) - M^2 A_{\mu \nu}A^{\mu \nu}\right] \nonumber \\
& & \!+\bar{\psi}(i\gamma^\lambda \partial_\lambda\! - \!m_\psi) \psi\! + \!\frac{1}{2}\left[(\partial_\lambda \varphi_{\rm ph}) (\partial^\lambda \varphi_{\rm ph})\! - \!m^2_\varphi \varphi^2_{\rm ph} \right]\! - \!\frac{1}{2}\sqrt{f}\left(F_{\mu \nu}\varphi\! + \!\zeta \bar\psi \sigma_{\mu\,\nu} \psi \right) A^{\mu \nu}, 
\end{eqnarray}

\vspace{-0.1cm}

\ni where $F_{\mu\,\nu} = \partial_\mu A_\nu - \partial_\nu A_\mu $, while $m_\psi$  and $m_\varphi$ are the sterino and steron masses, and $j_\mu$ denotes the \SM current. Here, $\varphi \equiv {<\!\!\varphi\!\!>_{\rm vac}} +\varphi_{\rm ph}$. 

It should be pointed out that the interaction $-(\sqrt{f}/2) F_{\mu\,\nu}\, \varphi\, A^{\mu\,\nu}$ of $\,F_{\mu\,\nu}\,$ with $\varphi \equiv  {<\!\!\varphi\!\!>_{\rm vac}} + \varphi_{\rm ph}$ and $A_{\mu\,\nu}$, appearing in the last term of Lagrangian (4), violates explicitly (in addition to the spontaneous breaking through the \SM Higgs mechanism) the electroweak symmetry $SU(2)\times U(1)$ in the presence of $\varphi$ and $A_{\mu\,\nu}$, because of the electromagnetic field $A_\mu = \sin \theta_{\rm w} W^0_\mu + \cos \theta_{\rm w} B_\mu$ which is not an $SU(2)\times U(1)$ scalar (in contrast to $\varphi$ and $A_{\mu\,\nu}$){\footnote {One can avoid such an explicit violation of the electroweak symmetry by the interaction $-(\sqrt{f}/2) F_{\mu\,\nu} \varphi A^{\mu\,\nu}$, if one omits this coupling in the last term of the Lagrangian (4). In that case, however, the photonic portal vanishes, as then the field equations (5) are reduced to $\partial_\nu F^{\mu\,\nu} = -j^\mu $ and $(\Box - M^2)A_{\mu\,\nu} = - \sqrt{f} \zeta \bar\psi \sigma_{\mu\,\nu} \psi $, and so are decoupled. Thus, the photonic portal is excluded by the (strict) electroweak symmetry and {\it vice versa} (somehow, they are dual or complementary to each other).}}. Part of this term, $-(\sqrt{f}{<\!\!\varphi\!\!>_{\rm vac}}/2) F_{\mu\,\nu} A^{\mu\,\nu}$, mixes photons with quanta of the field $A_{\mu\,\nu}$, making the electromagnetic field  a source of the field $A_{\mu\,\nu}$ and {\it vice versa} (see the coming Eqs.  (5), the second of which shows how the electromagnetic field $F_{\mu\,\nu}$ becomes a source of the field $A_{\mu\,\nu}$ when the constant ${<\!\!\varphi\!\!>_{\rm vac}} \neq 0$ stands in place of the field $\varphi$ on the rhs). The situation realized here in which the physical electromagnetic field may be a source of another dynamical field is new and astounding. If ${<\!\!\varphi\!\!>_{\rm vac}}$ were 0, such a situation would not occur, whereas the additional violation of electroweak symmetry by $F_{\mu\,\nu}$ still would be actual in the presence of $\varphi$ and $A_{\mu\,\nu}$. Thus, the mixing of $F_{\mu\,\nu}$ and $A_{\mu\,\nu}$ is a spontaneous phenomenon, being related to ${<\!\!\varphi\!\!>_{\rm vac}} \neq 0$.

The Lagrangian (4) implies two field equations that may be called {\it supplemented Maxwell's equations}:

\begin{equation}
\partial_\nu F^{\mu\,\nu} = -j^\mu - \delta j^\mu \;,\; (\Box - M^2)A_{\mu\,\nu} = - \sqrt{f}(F_{\mu\,\nu} \varphi   + \zeta \bar\psi \sigma_{\mu\,\nu} \psi)\;, 
\end{equation}

\ni where $F_{\mu\,\nu} = \partial_\mu A_\nu - \partial_\nu A_\mu$ and 

\begin{equation} 
\delta j^\mu \equiv \sqrt{f}\partial_\nu (\varphi A^{\mu\,\nu}) \equiv \partial_\nu \delta F^{\mu\,\nu} \;\;,\;\; \delta F^{\mu\,\nu} \equiv \sqrt{f} \varphi A^{\mu\,\nu}\;
\end{equation}

\ni ($\Box \equiv -\partial_\lambda \partial^\lambda$). Hence, 

\begin{equation} 
\partial_\mu \delta j^\mu \equiv 0.
\end{equation}

\ni If $-\Box$ can be neglected in the second Eq. (5) {\it versus} $M^2$, then from this equation we obtain 

\begin{equation}
A_{\mu\,\nu} \simeq  \frac{\sqrt{f}}{M^2}(F_{\mu\,\nu} \varphi   + \zeta \bar\psi \sigma_{\mu\,\nu} \psi) 
\end{equation}

\ni and  so,

\begin{equation}
\delta j^\mu \simeq \frac{f}{M^2}\partial_\nu \left[\varphi (F^{\mu\,\nu}\varphi + \zeta \bar\psi \sigma_{\mu\,\nu} \psi)\right] \;\;,\;\; \delta F^{\mu\,\nu} \simeq \frac{f}{M^2} \varphi (F^{\mu\,\nu}\varphi + \zeta \bar\psi \sigma_{\mu\,\nu} \psi)\,. 
\end{equation}

\ni Note that the first Eq. (5) may be rewritten also as

\begin{equation} 
\partial_\nu (F^{\mu\,\nu} + \delta F^{\mu\,\nu}) = -j^\mu \,,
\end{equation}

\ni where the field combination $F^{\mu\,\nu} + \delta F^{\mu\,\nu} = F^{\mu\,\nu} + \sqrt{f} \,\varphi \,A^{\mu\,\nu}$ together with the second Eq. (5) serves to supplement Maxwell's equations. Since $\partial_\mu \delta j^\mu \equiv 0$ holds identically (Eq. (7)), whereas $\partial_\mu j^\mu = 0$ is a dynamical equation, the correction $\delta j^\mu $ to the \SM electric current $j^\mu$ from the sterile world gets the {\it quasi-magnetic structure} (see also Eqs. (11) and (12) later on). 

Notice that eliminating $A_{\mu\,\nu}$  from the last term of  Lagrangian (4) by means of the approximate Eq. (8) one must divide the result by 2 to avoid the double-counting in such an operation leading to a quadratic form in the Lagrangian. Then, this term gives approximately the effective interaction (3).

Due to the nonzero vacuum expectation value ${<\!\!\varphi\!\!>_{\rm vac}} \neq 0$ of the steron field, we can see from the effective interaction (3) and the identity $\varphi \equiv {<\!\!\varphi\!\!>_{\rm vac}} +\varphi_{\rm ph}$ that sterinos, though they are sterile, display in the approximation (8) the effective {\it quasi-magnetic interaction}

\begin{equation} 
-\mu_\psi  \,\bar\psi \,\sigma_{\mu\,\nu} \,\psi\, F^{\mu\,\nu}
\end{equation}

\ni proportional to the sterino {\it quasi-magnetic moment}

\begin{equation} 
\mu_\psi \equiv \frac{f \zeta <\!\!\varphi\!\!>_{\rm vac}}{2M^2}
\end{equation}

\ni spontanously generated by the vacuum expectation value ${<\!\!\varphi\!\!>_{\rm vac}} \neq 0$ of steron field. Thus, the cold dark matter consisting of sterinos, whose mass is also spontaneously generated by  $
{<\!\!\varphi\!\!>_{\rm vac}} \neq 0$, is {\it quasi-magnetic} in the sense that it can interact effectively with cosmic as well as laboratory magnetic fields. In particular, it can be polarized in external magnetic fields. Such a spontaneous {\it quasi-magnetism} of cold dark matter is a charateristic feature of our model, where the photonic portal is open to the sterile world of sterinos and sterons.

Beside extended Maxwell's equations (5), the Lagrangian (4) implies also the Dirac and Klein-Gordon equations for sterino and steron fields $\psi$ and $\varphi_{\rm ph}$ interacting with $F_{\mu\,\nu}$ and $A_{\mu\,\nu}$, namely

\begin{equation}
(i \gamma^\mu \partial_\mu - \frac{1}{2} \sqrt{f} \zeta \sigma_{\mu\,\nu} A^{\mu\,\nu} - m_\psi) \psi = 0
\end{equation}

\ni and
 
\begin{equation}
(\Box - m^2_\varphi) \varphi_{\rm ph} =  \frac{1}{2} \sqrt{f} F_{\mu\,\nu} A^{\mu\,\nu}\,.
\end{equation}

\ni If $-\Box$ in the second Eq. (5) is negligible {\it versus} $M^2$, then in the Dirac equation (13) we can put $A^{\mu\,\nu} \simeq (\sqrt{f}/M^2) (F^{\mu\,\nu} \varphi + \zeta\,\bar{\psi}\,\sigma^{\mu\,\nu}\,\psi)$ (Eq. (8)), what --- with $\varphi \equiv {<\!\!\varphi\!\!>_{\rm vac}} +\varphi_{\rm ph}$  --- leads in this equation to the effective quasi-magnetic coupling $(f \zeta<\!\!\varphi\!\!>_{\rm vac}/2M^2) \sigma_{\mu\,\nu} F^{\mu\,\nu} \psi \equiv \mu_\psi \sigma_{\mu\,\nu} F^{\mu\,\nu}\psi$ (see Eq. (12)), spontaneously generated by ${<\!\!\varphi\!\!>_{\rm vac}} \neq 0$. 

Strictly speaking, for the derivation of Eqs. (5) as well as (13) and (14) the total Lagrangian (including the Lagrangian (4)) should be used. Then, the second Eq. (5) as well as Eqs. (13) and (14) get additional terms following from the rest of mass-generating couplings of $\varphi$ with itself and with $\psi$ and  $A_{\mu\,\nu}$. Here, they are suppressed for simplicity. The total Lagrangian implies also the rest of \SM field equations with the spontaneously broken electroweak symmetry.

Finally, we ought to make a tentative assumption about the magnitude of the coupling constant $f$. We decide to put   

\begin{equation}
f = e^2 \;\;{\rm with}\;\; e^2 \equiv 4\pi \alpha = \frac{1}{10.9} = 0.0917 
\end{equation}

\ni (for $\alpha = 1/137$). We also assume tentatively that the sterino mass $m_\psi$ is similar to the vacuum expectation value ${<\!\!\varphi\!\!>_{\rm vac}} \neq 0$  and to the mass scale $M$:
 
\begin{equation}
m_\psi \sim {<\!\!\varphi\!\!>_{\rm vac}} \sim M\,.
\end{equation}

\ni To the question of steron mass $m_\varphi$ we will turn later on.

In the case of assumption (15), our supplemented Maxwell's equations (5), unifying the \SM electrodynamics with the dynamics of cold dark matter, take the form 

\begin{equation}
\partial_\nu (F^{\mu\,\nu} + e \,\varphi\, A^{\mu\,\nu}) = -j^\mu \;,\; (\Box - M^2)A_{\mu\,\nu} = - e\, (F_{\mu\,\nu} \,\varphi   + \zeta \,\bar\psi \,\sigma_{\mu\,\nu} \,\psi)\;,  
\end{equation}

\ni where $F_{\mu\,\nu} = \partial_\mu A_\nu - \partial_\nu A_\mu$. Now,

\begin{equation}
\delta j^\mu \equiv e \,\partial_\nu(\varphi A^{\mu\,\nu}) \simeq \frac{e^2}{M^2} \partial_\nu \left[ \varphi (F^{\mu\,\nu} \varphi + \zeta \,\bar\psi\, \sigma^{\mu\,\nu} \,\psi ) \right]
\end{equation}

\ni and

\begin{equation}
\delta F^{\mu\,\nu} \equiv e \,\varphi\, A^{\mu\,\nu} \simeq \frac{e^2}{M^2} \varphi (F^{\mu\,\nu}\, \varphi + \zeta \,\bar\psi \,\sigma^{\mu\,\nu} \,\psi ) \,.
\end{equation}

\ni Of course, $j^\mu \propto e$ for the \SM electric current. Then, jointly with the assumption (16), we obtain for the sterino quasi-magnetic moment (12) the tentative estimation

\begin{equation}
\mu_\psi \sim \frac{e^2 \zeta}{2m_\psi} \,.
\end{equation}

\ni Here, the rhs is formally equal to the magneton of a fictitious particle carrying the electric charge $e^2 \zeta $ and having the sterino mass.  

\vspace{0.3cm}

\ni {\bf 3. Thermal freeze-out of sterinos}

\vspace{0.3cm} 

As is well known, the WAMP experimental estimate for the relic dark matter abundance [4] is

\begin{equation}
\Omega_{\rm DM}\,  h^2 \simeq 0.1 \,.
\end{equation}

\ni On the other hand, the thermal freeze-out mechanism in the case of weakly interacting massive particles (WIMPs) as candidates for the cold dark matter gives the following order-of-magnitude theoretical estimation [4]:

\begin{equation}
\Omega_{\rm DM}  \,h^2 \simeq \frac{3\times 10^{-27}\,{\rm cm}^3 {\rm s}^{-1}}{<\!\!\sigma_{\rm ann} v_{\rm DM}\!\!>}\,.
\end{equation}

\ni Hence, the experimental value for the thermal average of total annihilation cross-section (times the relative velocity) for a pair of cold dark matter particles can be estimated as

\begin{equation}
<\sigma_{\rm ann} v_{\rm {DM}}> \simeq 3\times 10^{-26} {\rm cm}^3{\rm s}^{-1} \simeq {\rm pb}\,  \simeq \frac{8}{\pi} \frac{10^{-3}}{{\rm TeV}^2}
\end{equation}

\vspace{0.3 cm}

\ni in the units where $c = 1$ and $\hbar c = 1$.

This thermal freeze-out value of $<\sigma_{\rm ann} v_{\rm {DM}}>$ happens to be consistent with the typical size of weak-interaction cross-sections, what provides a strong numerical argument for neutralinos of the supersymmetric extension of \SM as candidates for WIMPs (as well as for the thermal mechanism of their decoupling in the early Universe). 

As we will see, also sterinos with an appropriate mass can be considered as possible candidates for WIMP cold dark matter. To this end, let us calculate the cross-sections in the simplest annihilation channel for a sterino-antisterino pair,

\begin{equation}
({\rm antisterino})({\rm sterino}) \rightarrow \gamma\, ({\rm steron}) \,, 
\end{equation}

\ni and in the channel a bit more complicated
 
\begin{equation}
({\rm antisterino})({\rm sterino}) \rightarrow e^+e^- \,.
\end{equation}

\ni Applying the interaction (following from the Lagrangian (4))

\begin{equation}
-\frac{1} {4}\; \frac{2f \zeta}{M^2} F_{\mu\,\nu} \varphi_{\rm ph} (\bar\psi \sigma^{\mu\,\nu} \psi ) 
\end{equation}

\vspace{0.2cm}

\ni in the first case, and the interaction

\begin{equation}
-\frac{1} {4}\; \frac{2f \zeta}{M^2} F_{\mu\,\nu} {<\!\!\varphi\!\!>_{\rm vac}} (\bar\psi \sigma^{\mu\,\nu} \psi ) + e\,\bar{\psi}_e\,\gamma^\mu \,\psi_e A_\mu
\end{equation}

\ni ($e = |e|$) in the second, we obtain in the sterino-antisterino centre-of-mass frame the following formulae:

\begin{equation}
\sigma(\bar{\psi} \psi \rightarrow \gamma \varphi_{\rm ph}) 2v_\psi = \frac{1}{6\pi} \left(\frac{f \zeta}{M^2}\right)^{\!\!2}\left(E^2_\psi + 2m^2_\psi\right) \left(1 -   \frac{m^2_\varphi}{4 E^2_\psi}\right) 
\end{equation}

\ni and

\begin{equation}
\sigma(\bar{\psi} \psi \rightarrow e^+ e^-) 2v_\psi = \frac{1}{12\pi} \left(\frac{e f \zeta<\!\!\varphi\!\!>_{\rm vac}}{M^2}\right)^{\!\!2}\frac{E^2_\psi + 2m^2_\psi}{E^2_\psi} \,,
\end{equation}

\ni where in the second channel the electron mass is neglected ($E_e = E_\psi \geq m_\psi  \gg m_e$). Hence, their small ratio is 

\begin{equation}
\frac{\sigma(\bar{\psi} \psi \rightarrow e^+ e^-)}{\sigma(\bar{\psi} \psi \rightarrow \gamma \varphi_{\rm ph})}
\leq \frac{e^2}{2} <\!\!\varphi\!\!>^2_{\rm vac} \left(m^2_\psi - \frac{m^2_\varphi}{4} \right)^{-1} \sim  \frac{e^2}{2} \left(1 - \frac{m^2_\varphi}{4m^2_\psi } \right)^{-1} \,,
\end{equation}

\ni since $E_\psi \geq m_\psi \sim <\!\!\varphi\!\!>_{\rm vac}$ tentatively (Eq. (16)), and further 

\vspace{0.1cm}

\begin{equation}
\sim \frac{1}{16.4}\;\;{\rm or}\;\; \frac{1}{21.8}
\end{equation}

\ni if in addition

\begin{equation}
m_\varphi^2 \sim m^2_\psi \;\;{\rm or}\;\; m_\varphi^2 \ll 4m^2_\psi \;,
\end{equation}

\ni respectively. Thus, putting approximately

\begin{equation}
\sigma_{\rm ann} v_{\rm {DM}} \simeq \sigma(\bar{\psi} \psi \rightarrow \gamma \varphi_{\rm ph}) 2v_\psi \,,
\end{equation}

\ni we get the following formula in the approximation of $E_\psi \simeq m_\psi$ ({\it i.e.}, $\vec{p}^{\,2}_\psi/m^2_\psi\ll 1$):

\begin{equation}
\sigma_{\rm ann} v_{\rm {DM}} \simeq \frac{1}{2\pi} \left(\frac{f \zeta}{M^2}\right)^{\!\!2} m^2_\psi \left(1 -   \frac{m^2_\varphi}{4m^2_\psi}\right) \,.
\end{equation}

\ni If the additional conditions (32) hold, the formula (34) implies that

\begin{equation}
\sigma_{\rm ann} v_{\rm {DM}} \sim \frac{1}{2\pi} \left(\frac{3}{4} \;\;{\rm or} \;\; 1\right)\left(\frac{f \zeta}{M^2} \right)^{\!\!2} m^2_\psi \,,
\end{equation}

\ni respectively. Since $f = e^2$ and $m_\psi \sim M$ tentatively (Eqs. (15) and (16)), we get respectively the estimation

\begin{equation}
\sigma_{\rm ann} v_{\rm {DM}} \simeq \frac{1}{2\pi} \left(\frac{3}{4} \;\;{\rm or} \;\; 1\right) \left(\frac{e^2 \zeta}{m_\psi} \right)^{\!\!2} \,.
\end{equation}

\ni Here, $\!<\!\!\sigma_{\rm ann} v_{\rm {DM}}\!\!>\, \simeq \sigma_{\rm ann} v_{\rm {DM}}$, because  $\sigma_{\rm ann} v_{\rm {DM}} $ as given in Eq. (35) is independent of $ v_{\rm DM} = 2v_\psi$.

Comparing Eqs. (23) and (36), we can estimate the sterino mass $m_\psi$ in terms of the parameter $\zeta$: 

\begin{equation}
m_\psi \sim \left(\frac{\sqrt{3}}{2} \;\;{\rm or} \;\; 1\right)\frac{e^2 \zeta}{4}\;{\rm TeV} = \left(0.6 \;\;{\rm or} \;\; 0.7\right)\,\zeta\;{\rm TeV}\,,
\end{equation}

\ni where the conditions (32) are applied, respectively. We can see that the thermal freeze-out value of sterino mass $m_\psi$  is pretty stable for steron mass in the range $0 < m_\varphi \stackrel{<}{\sim} m_\psi$. For such values of $m_\varphi$, sterinos with the mass $m_\psi \sim (0.6 - 0.7)\zeta$ TeV are possible candidates for the cold dark matter decoupled thermally in the early Universe. This conclusion is obtained under the tentative assumptions of $f = e^2$ and $m_\psi \sim  <\!\!\varphi\!\!>_{\rm vac} \sim M$. The natural value of the parameter $\zeta $ is here $\zeta \sim 1$, as then in the effective interaction (3) the proportion of its three coupling constants is the simplest one, being $f : 2f \zeta : f \zeta^2 \sim 1 : 2 : 1$ like for the coefficients in the quadratic form $(a+b)^2 = a^2 + 2ab + b^2$.

In contrast to sterinos, sterons are unstable. Their simplest decay channel is

\begin{equation}
({\rm steron}) \rightarrow \gamma\, \gamma
\end{equation}

\ni with the following decay rate at rest: 

\begin{equation}
\Gamma(\varphi_{\rm ph} \rightarrow {\gamma \gamma}) = \frac{1}{128\pi} \left(
\frac{f<\!\!\varphi\!\!>_{\rm vac}}{M^2} \right)^{\!\!2} \,m^3_\varphi \sim \frac{1}{128\pi} \left(
\frac{e^2}{m_\psi} \right)^{\!\!2} \,m^3_\varphi \sim \frac{e^4}{128\pi} \left( m_\psi \;{\rm or}\; 
\frac{m^3_\varphi}{m^2_\psi}\right) \;, 
\end{equation}

\ni where in the last step the conditions (32) hold, respectively, and our tentative assumptions of $f = e^2$ and $m_\psi \sim\, <\!\!\varphi\!\!>_{\rm vac}\,\sim M$ are applied. Here, $e^4/128 \pi = \pi \alpha^2/8 = 2.09\times 10^{-5}$. In the case of $ m^2_\varphi \sim m^2_\psi$ , the formula (39) gives

\begin{equation}
\Gamma(\varphi_{\rm ph} \rightarrow {\gamma \gamma}) \sim 10\, \zeta \;{\rm MeV} = 2\times 10^{22} \zeta\, {\rm s}^{-1}
\end{equation}

\ni ($\hbar =1$), thus the life-time becomes $\tau(\varphi_{\rm ph} \rightarrow {\gamma \gamma})  \equiv \Gamma^{-1}(\varphi_{\rm ph} \rightarrow {\gamma \gamma})  \sim 5\times 10^{-23} \zeta^{-1}$ s . In the case of $ m_\varphi^2 \ll 4m^2_\psi$, putting {\it e.g.} $m_\varphi = \zeta$ MeV, we obtain from the formula (39)  

\begin{equation}
\Gamma(\varphi_{\rm ph} \rightarrow {\gamma \gamma}) \sim 4\times 10^{-17} \zeta \;{\rm MeV} = 6\times 10^{4} \,\zeta\; {\rm s}^{-1}
\end{equation}

\ni and so, $\tau(\varphi_{\rm ph} \rightarrow {\gamma \gamma})  \equiv \Gamma^{-1}(\varphi_{\rm ph} \rightarrow {\gamma \gamma})  \sim 2\times 10^{-5} \zeta^{-1}$ s . Here, the natural value of $\zeta$ is $\zeta \sim 1$. One may ask the question, what is the lower bound for the mass of steron unstable at the Universe scale {\it i.e.}, such that $\Gamma(\varphi_{\rm ph} \rightarrow {\gamma \gamma}) \stackrel {>}{\sim}\; {\rm(age\; of\; the\; Universe)}^{-1} = (13.7\;{\rm Gyr})^{-1} = (4.3\times 10^{17}\; {\rm s})^{-1}$ [3]. Putting $m^2_\varphi \ll 4 m^2_\psi$ and $m_\psi \sim 0.7\,\zeta $ TeV, one obtains then from Eq. (39) that

\begin{equation} 
m_\varphi \stackrel {>}{\sim}\, 3\times 10^{-2} \zeta^{2/3} \;{\rm eV} \,.
\end{equation}

Due to the quasi-magnetism of sterinos, they could be produced in high energy collisions of \SM particles {\it via} virtual photons, as in the simplest process of this kind 

\begin{equation} 
e^+e^- \rightarrow \gamma^* \rightarrow ({\rm antisterino})({\rm sterino}) \,,
\end{equation}

\ni inverse to the process (25). In this case, the centre-of-mass threshold energy is $2 m_\psi \sim\,(1.2-1.4)\,\zeta$~TeV (if our estimation of $m_\psi$ is valid). Applying the interaction (27), we obtain in the centre-of-mass frame the following cross-section:

\begin{eqnarray}
\sigma(e^+e^- \!\rightarrow \bar{\psi} \psi) 2v_e & = & \frac{1}{12\pi}\!  \left(\frac{e f \zeta<\!\!\varphi\!\!>_{\rm vac}}{M^2}\right)^{\!\!2}\!\!\left(1+\frac{m^2_\psi}{2E^2_e}\right) \sqrt{1-\frac{m^2_\psi}{E^2_e}} \nonumber \\ & \sim & \frac{1}{12\pi}  \left(\frac{e^3 \zeta}{m_\psi}\right)^{\!\!2}\!\!\left(1+\frac{m^2_\psi}{2E^2_e}\right) \sqrt{1-\frac{m^2_\psi}{E^2_e}} \,,
\end{eqnarray}

\ni where the electron mass is neglected ($E_e = E_\psi \stackrel{>}{\sim} (0.6-0.7) \zeta$ TeV $\gg m^2_e$). In the second step, our tentative assumptions $f = e^2$ and $m_\psi \sim  \,<\!\!\varphi\!\!>_{\rm vac}\, \sim M$ are used. Here, $e^6/12\pi =16\pi^2 \alpha^3/3 = 2.05\times 10^{-5}$.

\vspace{0.3cm}

\ni {\bf 4. Conclusions}

\vspace{0.3cm}

We have proposed new simple field equations (5) that unify Maxwell's equations of the \SM electrodynamics (after the electroweak symmetry is spontaneously broken) with the dynamics of cold dark matter. The cold dark matter consists in our work of sterile spin-1/2 Dirac fermions (sterinos) whose mass is spontaneously generated by the nonzero vacuum expectation value of a field of sterile spin-0 bosons (sterons), while their interactions are mediated by quanta of a sterile antisymmetric tensor field with a large mass. The new field equations include a specific weak coupling of the \SM electromagnetic field to the sterile fields responsible for cold dark matter. It has a spontaneous quasi-magnetic structure and acts independently of gravity. Such a coupling opens a photonic portal between the \SM world and the sterile world consisting of sterinos, sterons and quanta of a sterile mediating field.

The thermal freeze-out mechanism in  the early Universe works well for sterinos with the mass $(0.6 - 0.7) \,\zeta$ TeV (here, the value $\zeta \sim 1$ is natural). This estimation is obtained under some tentative assumptions about the strength of the specific weak coupling and its mass scale. Thus, sterinos turn out to be possible candidates for WIMP cold dark matter decoupled thermally in the early Universe.
 
\vspace{1.0cm}

{\centerline{\bf References}}

\vspace{0.3cm}

{\everypar={\hangindent=0.65truecm}
\parindent=0pt\frenchspacing

{\everypar={\hangindent=0.65truecm}
\parindent=0pt\frenchspacing

~[1]~W.~Kr\'{o}likowski, arXiv: 0712.0505 [{\tt hep--ph}] (to appear in a more extended version in {\it Acta Phys. Polon.} {\bf B}); arXiv: 0803.2977v2 [{\tt hep--ph}]; arXiv: 0805.0675 [{\tt hep--ph}].

\vspace{0.2cm}

~[2]~ For some recent publications {\it cf.} J.~March-Russell, S.M.~West, D.~Cumberbath and D.~Hooper, arXiv: 0801.3440v2 [{\tt hep-ph}]; K.Y.~Lee, Y.G.~Kim and S.~Shin, arXiv: 0803.2932 [{\tt hep-ph}]; and references therein.

\vspace{0.2cm}

~[3]~Particle Data Group, {\it Review of Particle Physics, J. Phys}, {\bf G 33}, 1 (2006).

\vspace{0.2cm}

~[4]~For recent reviews {\it cf.} G. Bartone, D.~Hooper and J.~Silk, {\it Phys. Rept.} {\bf 405}, 279 (2005); M.~Taoso, G.~Bartone and A.~Masiero, arXiv: 0711.4996 [{\tt astro-ph}]; {\it cf.} also E.W.~Kolb and S.~Turner, {\it Early Universe} (Addison-Wesley, Reading, Mass., 1994); and K.~Griest and D.~Seckel, {\it Phys. Rev.} {\bf D 43}, 3191 (1991). 

\vfill\eject

\end{document}